\documentclass[prl,twocolumn,floatfix,showpacs,superscriptaddress]{revtex4-1}

\usepackage{amsmath}
\usepackage{amssymb}
\usepackage{amsfonts}
\usepackage[]{graphicx}
\usepackage{units}
\usepackage{bm}
\usepackage{xspace}
\usepackage[usenames]{color}

\newcommand{\graySn}{\mbox{$ \alpha $-Sn}\xspace}  
\newcommand{\GP}{$\Gamma$ point\xspace}   
\newcommand{\celcius}{$ \,^{\circ}\mathrm{C} $\xspace}     
\newcommand{\EF}{$\mathrm{\textit{E}_{F}}$\xspace}

\newcommand{\GSM}{$ \mathrm{\Gamma_{7}^{-}} $\xspace} 
\newcommand{\GSP}{$ \mathrm{\Gamma_{7}^{+}} $\xspace} 
\newcommand{\GEP}{$ \mathrm{\Gamma_{8}^{+}} $\xspace} 

\begin{document}


\title{Double Band Inversion in $ \alpha $-Sn: Appearance of Topological Surface States \\
and the Role of Orbital Composition}

\author{Victor A. Rogalev}
\affiliation{Physikalisches Institut und R{\"o}ntgen Center for Complex Material Systems, Universit{\"a}t W{\"u}rzburg, 97074 W{\"u}rzburg, Germany}

\author{Tom\'{a}\v{s} Rauch}%
\affiliation{Institute of Physics, Martin Luther University Halle-Wittenberg, 06099 Halle (Saale), Germany}

\author{Markus R. Scholz}
\affiliation{Physikalisches Institut und R{\"o}ntgen Center for Complex Material Systems, Universit{\"a}t W{\"u}rzburg, 97074 W{\"u}rzburg, Germany}

\author{Felix Reis}
\affiliation{Physikalisches Institut und R{\"o}ntgen Center for Complex Material Systems, Universit{\"a}t W{\"u}rzburg, 97074 W{\"u}rzburg, Germany}

\author{Lenart Dudy}
\affiliation{Physikalisches Institut und R{\"o}ntgen Center for Complex Material Systems, Universit{\"a}t W{\"u}rzburg, 97074 W{\"u}rzburg, Germany}

\author{Andrzej Fleszar}
\affiliation{Institut f{\"u}r Theoretische Physik und Astronomie, Universit{\"a}t W{\"u}rzburg, 97074 W{\"u}rzburg, Germany}

\author{Marius-Adrian Husanu}
\affiliation{Swiss Light Source, Paul Scherrer Institute, CH-5232 Villigen, Switzerland} %
\affiliation{National Institute of Materials Physics, 077125, Magurele, Romania} %

\author{Vladimir N. Strocov}
\affiliation{Swiss Light Source, Paul Scherrer Institute, CH-5232 Villigen, Switzerland} %

\author{J\"urgen Henk}%
\affiliation{Institute of Physics, Martin Luther University Halle-Wittenberg, 06099 Halle (Saale), Germany}

\author{Ingrid Mertig}%
\affiliation{Institute of Physics, Martin Luther University Halle-Wittenberg, 06099 Halle (Saale), Germany}
\affiliation{Max Planck Institute for Microstructure Physics, 06120 Halle (Saale), Germany} %

\author{J{\"o}rg Sch{\"a}fer}
\affiliation{Physikalisches Institut und R{\"o}ntgen Center for Complex Material Systems, Universit{\"a}t W{\"u}rzburg, 97074 W{\"u}rzburg, Germany}

\author{Ralph Claessen}
\affiliation{Physikalisches Institut und R{\"o}ntgen Center for Complex Material Systems, Universit{\"a}t W{\"u}rzburg, 97074 W{\"u}rzburg, Germany}


\date{\today}

\begin{abstract}

 The electronic structure of \graySn(001) thin films strained compressively in-plane was studied both experimentally and theoretically. A new topological surface state (TSS) located entirely within the gapless projected bulk bands is revealed by \textit{ab initio}-based tight-binding calculations as well as directly accessed by soft X-ray angle-resolved photoemission. The topological character of this state, which is a surface resonance, is confirmed by unravelling the band inversion and by calculating the topological invariants. In agreement with experiment, electronic structure calculations show the maximum density of states in the subsurface region, while the already established TSS near the Fermi level is strongly localized at the surface. Such varied behavior is explained by the differences in orbital composition between the specific TSS and its associated bulk states, respectively. This provides an orbital protection mechanism for topological states against mixing with the background of bulk bands.

\end{abstract}


\maketitle

\begin{figure} [!htb]
	\includegraphics[clip, trim=1.5cm 2.5cm 1.5cm 1cm, width=1\linewidth]{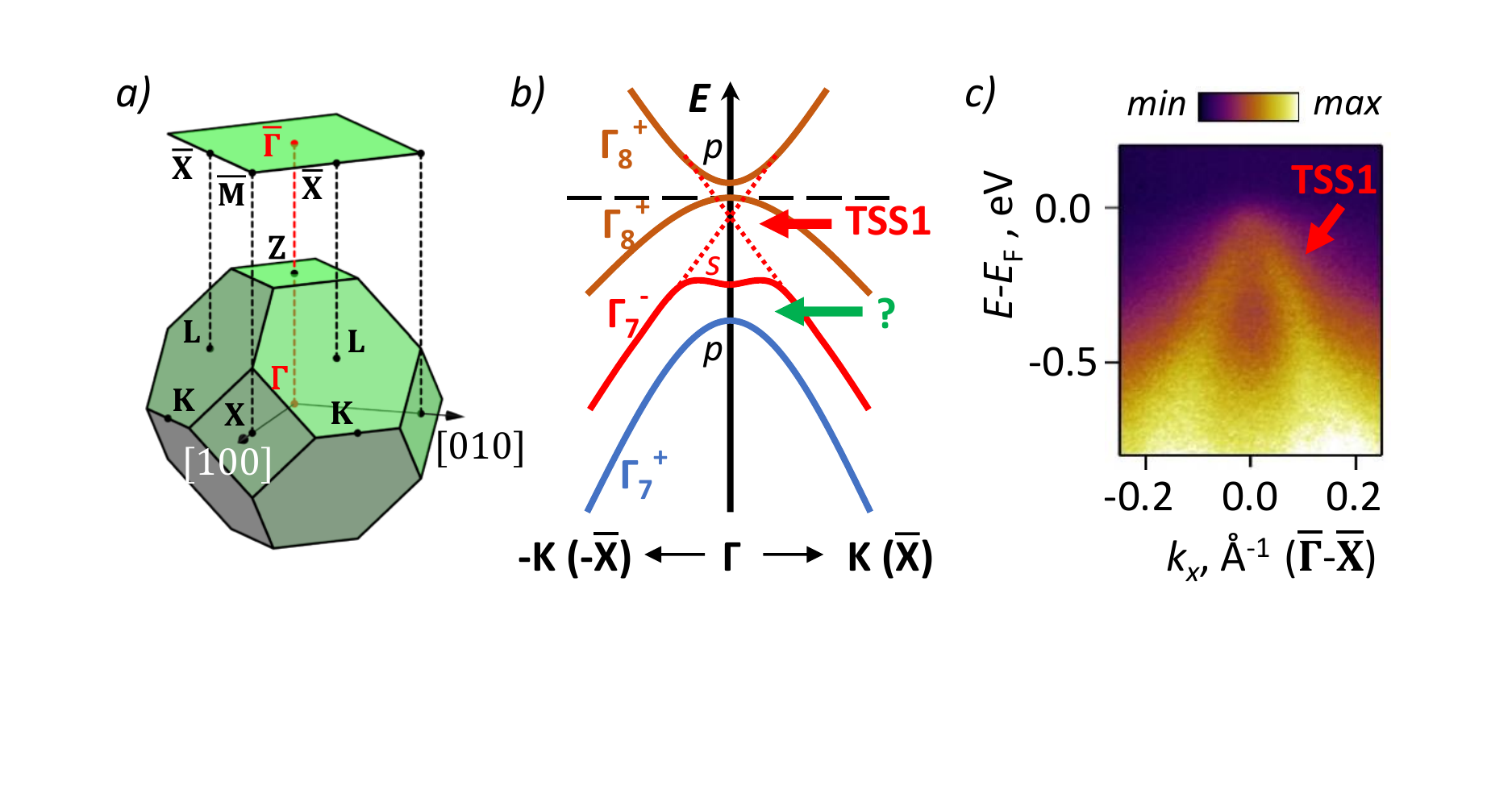}
	\caption{(Color online) Brillouin zone and electronic structure of strained \graySn(001). a) Bulk and surface BZ\@. b) Sketch of the band structure along $\mathrm{\Gamma K X} $ ($\overline{\Gamma}\, \overline{\mathrm{X}}$) assuming in-plane (001) strain and SOC\@. Band parities are encoded by $+$ or $-$ superscripts, while dominant orbital compositions near the \GP (\textit{p} or \textit{s}) identify two band inversions (indicated by red and green arrows). c) ARPES data taken with the He-I line ($h\nu = \unit[21.2]{eV}$) show TSS1 near \EF.}
	\label{Intro}
\end{figure}

Since the theoretical prediction of topological insulators (TIs) \cite{Kane2005,Kane2005a}, many materials were found to belong to this new class of solids \cite{Hasan2010,Bansil2016}. The interest in these materials is based on the idea to exploit and to manipulate spin-polarized metallic surface states with forbidden backscattering due to time-reversal symmetry. These so-called topological surface states (TSSs) are induced at an interface between systems that belong to different topological phases. 
The topologically non-trivial low temperature phase of Sn (\graySn) is a particularly interesting example for such a system.
In contrast to other TIs, like the $ \mathrm{Bi_{2}X_{3}} $ compounds, in which the inverted bands are separated by a global band gap \cite{Zhang2009}, unstrained \graySn is a semimetal  \cite{Brudevoll1993,Pedersen2010} with diamond structure and the Fermi energy (\EF) pinned to the \GEP level (Fig.~\ref{Intro}). Spin-orbit coupling (SOC) shifts the $p$-derived split-off \GSP band below the $s$-derived \GSM band which in turn is situated below the four-fold degenerate $p$-derived \GEP band \cite{Barfuss2013c,Kufner2013a}. The latter degeneracy can be lifted by applying strain [Fig.~\ref{Intro}(b)]. The key feature of topological materials --- the band inversion --- occurs in \graySn between the \GEP and the \GSM levels at the \GP of the Brillouin zone (BZ). Importantly, yet another band inversion exists at the \GP of the BZ within the occupied electronic structure of \graySn: between \GSM and \GSP bands, which has gone unnoticed in previous studies.

Although \graySn has been predicted to be topologically nontrivial \cite{Fu2007b}, neither the electronic structure nor the existence of surface states was discussed in detail until the experimental observation of a TSS which emerges between the \GSM  and \GEP levels and exhibits a Dirac point close to \EF \cite{Barfuss2013c,Ohtsubo2013b,Rojas-Sanchez2016}; we refer to this state as TSS1 from now on. Interestingly, although TSS1 is largely degenerate with the surface-projected bulk \GEP band, in angle-resolved photoelectron spectroscopy (ARPES) it is detected as a sharply defined dispersive peak [Fig.~\ref{Intro}(c)]. Moreover, a very recent experimental report on spin-to-charge conversion supposedly involving the TSS1 of \graySn \cite{Rojas-Sanchez2016} demonstrates a practical utility of this material for spintronics. The above results raise two questions: what is the mechanism that protects TSS1 from hybridization with the bulk states, and can we expect another TSS (TSS2) induced by the other band inversion?

Here we report on a new surface state in the occupied electronic structure of strained \graySn{}(001) films, explored in a combined theoretical and experimental approach. Electronic-structure calculations reveal that this state bridges the \GSM  and \GSP bands and is located in the subsurface region, thus penetrating deeper into the bulk than TSS1 \cite{Barfuss2013c,Ohtsubo2013b}. Bulk-sensitive Soft X-ray ARPES (SX-ARPES) provides a direct experimental evidence for this new subsurface state, which has been overlooked in previous low-photon-energy ARPES experiments with smaller probing depths. Analyzing the orbital characters of bulk and surface states, we show that the different decay character of two surface states is explained by hybridization with the corresponding bulk states. In particular, TSS1's orbital composition suppresses hybridization with surface-projected bulk states, while for the new surface state such hybridization is favored. Moreover, we review the topological properties of the bulk-band structure by calculating the $\mathcal{Z}_{2}$ invariant as well as the mirror Chern numbers. We find that the new `buried' surface state is topologically nontrivial despite being located further away from the topological phase discontinuity (interface TI/vacuum).

Thin films of \graySn were grown \textit{in situ} on InSb(001) substrates. Prior to deposition, substrates have been cleaned in a series of sputter-anneal cycles, while the surface quality was monitored by low-energy electron diffraction (see Ref.~\onlinecite{Suppl2016}). Sn was deposited on the substrates at room temperature, simultaneously  Te was deposited to $n$-dope the film and to provide a smooth growth \cite{Barfuss2013c}. The thickness of the \graySn film was at least $\approx \unit[6]{nm} $ (9 unit cells) as estimated from photoemission data \cite{Suppl2016}. SX-ARPES measurements have been performed at the ADRESS beamline of the Swiss Light Source. Details of the experimental setup are given in Ref.~\onlinecite{Strocov2014}. While bulk \graySn is stable only below $\approx 13.2 $\celcius \cite{Jayaraman1963}, thin films of \graySn epitaxially deposited on InSb(001) substrates are stable up to $\approx 200$\celcius \cite{Mason1992}, which guarantees successful growth at room temperature.

The InSb substrate induces compressive strain in \graySn due to the lattice mismatch of $\unit[0.13]{\%}$, the latter  causing a local band gap of $\approx \unit[30]{meV}$ at the \GP of the bulk Brillouin zone. Similar to HgTe \cite{Rauch2015,Ruan2016}, a global band gap in the entire reciprocal space is opened only by tensile in-plane  (001) strain, whereas for compressive in-plane (001) strain the local band gap around the bulk \GP closes along the $ \mathrm{\Gamma} $-Z direction \cite{Yang2014a,Suppl2016}. The resulting band crossing gives rise to a Dirac point in the bulk band structure. These properties suggest that compressively strained \graySn is a Dirac semimetal. We note, that the closing of band gap does not alter the topological character \cite{Barfuss2013c,Ohtsubo2013b} because the former occurs between bands whose associated states have even parity [Fig.~\ref{Intro}(b)].

To explore the bulk and surface electronic structure of \graySn we first performed \textit{ab initio}-based tight binding (TB) calculations for bulk and semi-infinite systems. The Slater-Koster parameters \cite{Slater1954,Suppl2016} for first- and second-nearest neighbors as well as the spin-orbit coupling strength were obtained by optimizing the TB band structure with respect to \textit{ab initio} data \cite{Brudevoll1993,Kufner2013a}. Special attention has been paid to reproducing the correct band ordering near the \GP.

\begin{figure} [!htb]
	\includegraphics[clip, trim=0cm 0.3cm 0.5cm 0.5cm, width=1\linewidth]{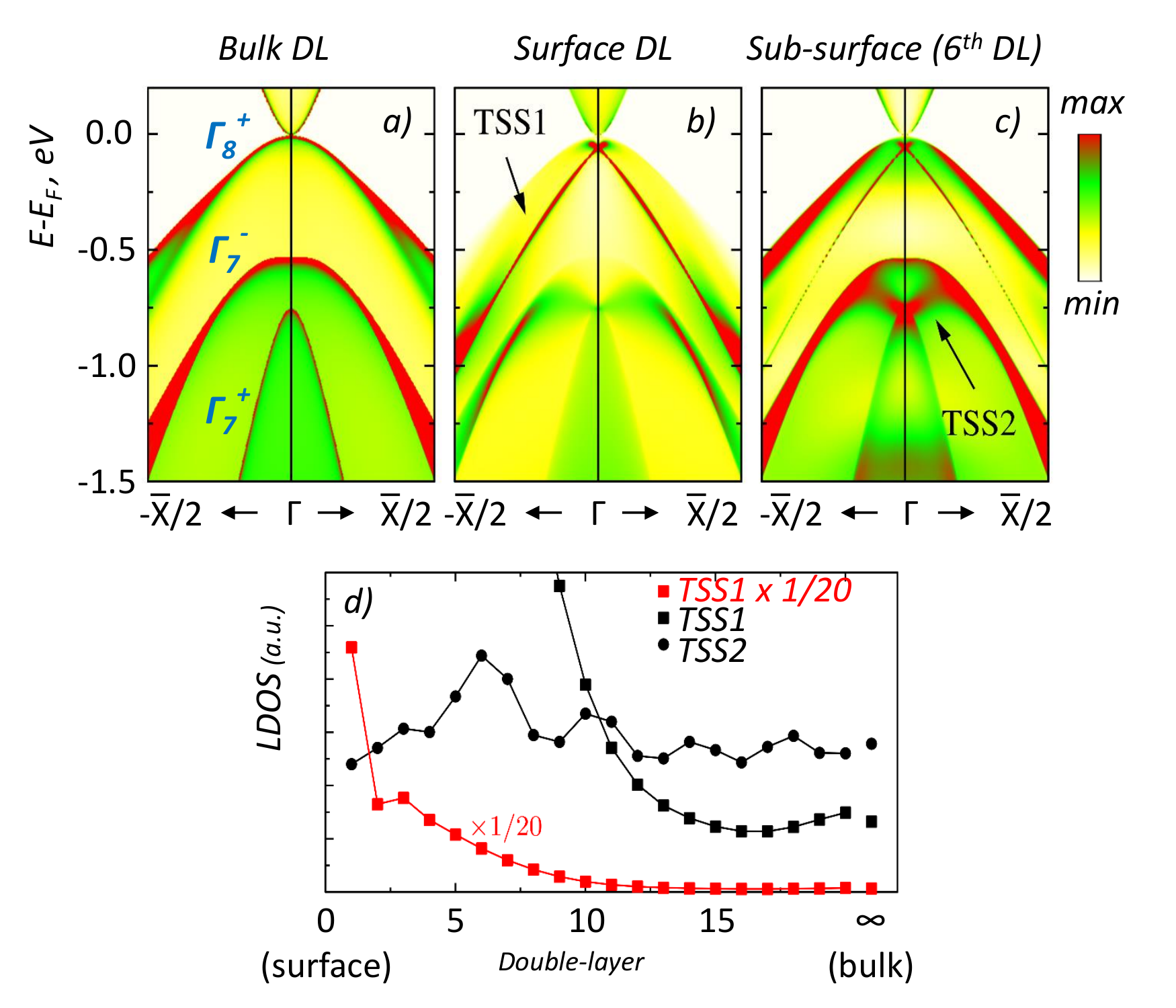}
	\caption{(Color online) Theoretical electronic structure of \graySn(001) obtained by TB calculations. The spectral density, given by color scale, is depicted for a) a bulk double-layer (DL), b) a surface DL, and c) the sixth subsurface DL. d) Layer-resolved spectral density near $\overline{\Gamma} $ for TSS1 (red and black squares) and TSS2 (black circles). While TSS1 is surface-localized, TSS2 is located in the subsurface region, with an LDOS maximum in the sixth DL\@. Red and black squares show identical but differently scaled data for TSS1.}
	\label{THEOR-1N}
\end{figure}

\begin{figure*} [!ht]
	\includegraphics[clip, trim=0cm 0.8cm 0cm 0cm, width=0.9\linewidth]{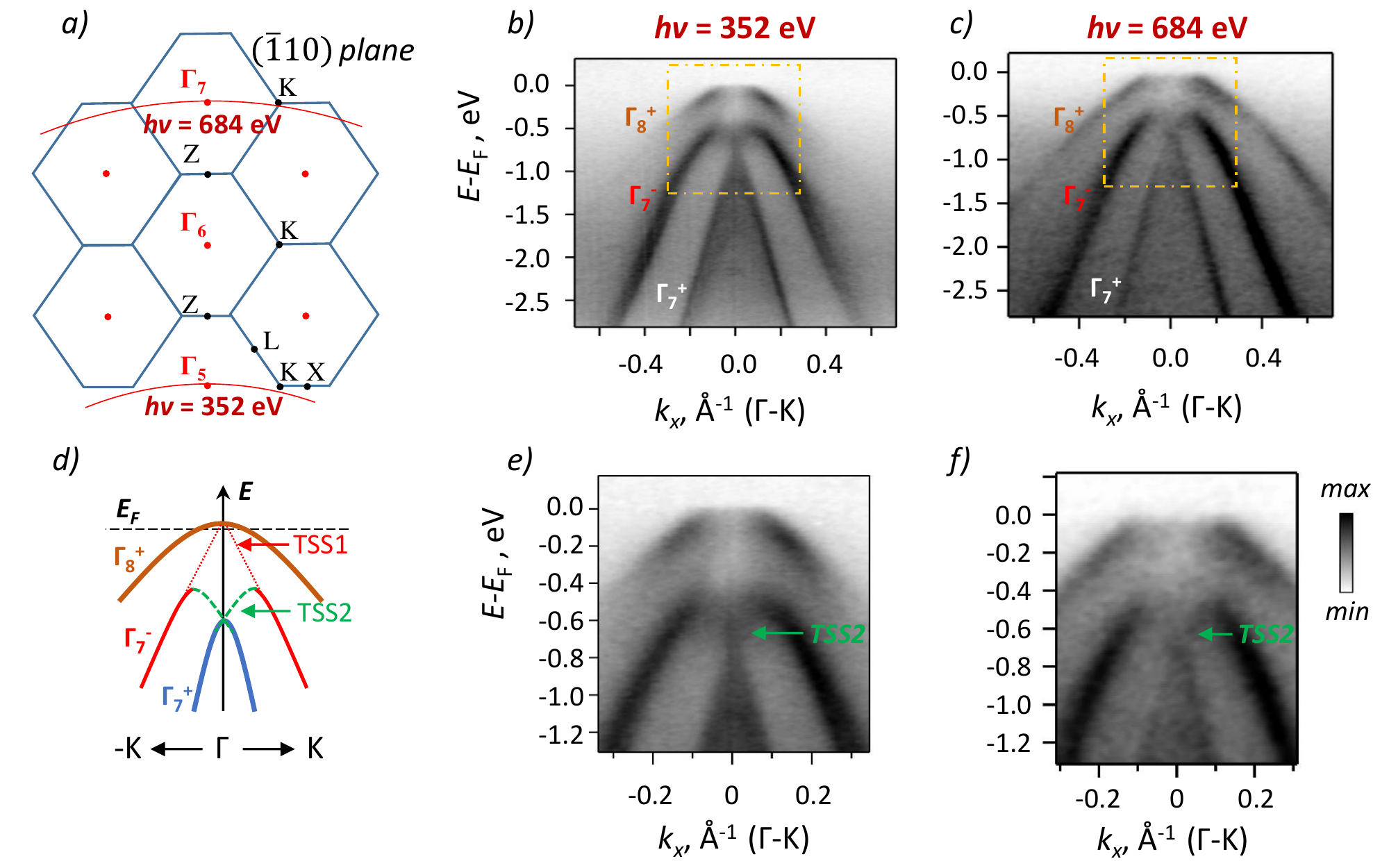}
	\caption{Experimental electronic structure of \graySn(001). a) Sketch of the bulk BZ in the  $\mathrm{(\overline{1}10)}$ mirror plane and ARPES measurement path for two photon energies. b) ARPES map $I(E, k_{x}$) taken for $ h\nu = \unit[352]{eV}$ and p-polarized light along $\mathrm{\Gamma K}$ ($\overline{\Gamma}\, \overline{\mathrm{X}}$). c) As b) but $h\nu = \unit[684]{eV}$ and for circular polarized light. d) Sketch of the observed dispersions based on the data shown in b) and c). e) and f) Zoom into regions shown with yellow frames on b) and c). Although the intensity of TSS1 is weak, bulk bands and TSS2 are clearly resolved. TSS1 becomes more visible in a second derivative of the ARPES map $I(E, k_{x})$ \cite{Suppl2016}.}
	\label{ARPES-1}
\end{figure*}

The TB results are presented in Fig.~\ref{THEOR-1N} (see also \cite{Suppl2016}), where we show a depth distribution of the electronic states' wavefunctions. When projected onto a Sn double layer (DL; half of the cubic unit cell) within the bulk [Fig.~\ref{THEOR-1N}(a)], the \GEP, \GSM and \GSP bands remain unconnected, in agreement with the bulk electronic structure \cite{Brudevoll1993}. On the contrary, when projected onto the DL near the surface [Figs.~\ref{THEOR-1N}(b) and ~\ref{THEOR-1N}(c)], TSS1 shows up as a consequence of the \GSM / \GEP band inversion, exhibiting the maximum spectral density in the topmost DL. Importantly, a new `connecting' spectral density located between the \GSM and \GSP bulk bands appears at $\approx \unit[-0.8]{eV}$ [Fig.~\ref{THEOR-1N}(c)]; we refer to this feature as TSS2 in the following. The spectral density of TSS2 is enhanced in the sub-surface region,  having an extended maximum at and around the sixth DL, as shown in Fig.~\ref{THEOR-1N}(d). Although the dispersive behavior may at first glance look like a Rashba split \GSM band, such an explanation can be ruled out, as its overall band width does not change from surface to bulk, whereas a Rashba effect would occur only at the surface and inevitably increase the width of the \GSM band. Based on our slab-calculations \cite{Suppl2016} a quantum well origin of this state can be excluded as well.

As pointed out by Barfuss \textit{et al.} \cite{Barfuss2013c}, spin-orbit coupling lifts the degeneracy of the \GSP and \GEP bands at \EF and pushes the \GSP band below the \GSM band, thereby introducing another band inversion. Therefore, the so far undiscovered TSS2 exists in an `inverted' SOC-induced band gap, which suggests a topological origin [Fig.~\ref{Intro}(b)].
This conjecture and, thus, the topological nontrivial band ordering are affirmed by calculations of topological invariants. Originally defined for insulating systems \cite{Fu2007}, the $\mathcal{Z}_{2}$ invariant has also been calculated for systems without a global band gap \cite{Fukui2007,Teo2008}. A necessary condition for such calculations is the presence of a finite energy difference at each wavevector $\vec{k}$ (i.\,e., a $\vec{k}$-dependent band gap). This is the case for \graySn in which the \GSP band is separated from the \GSM band in the entire bulk BZ, irrespectively whether moderate compressive or tensile strain is applied (strain less than $\unit[3]{\%}$ is assumed). The $\mathcal{Z}_{2}$ invariant of the bands below the \GSM band was calculated to $\left( \nu_{0};\nu_{1}\nu_{2}\nu_{3} \right) = \left(1;000\right)$ by using the Fu-Kane formula \cite{Fu2006} discretized by Fukui and Hatsugai \cite{Fukui2007} and by tracing maximally localized Wannier functions \cite{Yu2011}. Moreover, the lattice is invariant upon reflection at the (110) and the $(1\overline{1}0)$ mirror planes; the respective mirror Chern numbers are $n_{\mathrm{M}} = -1$ for both planes (further details in the Supplemental Material \cite{Suppl2016}). Both topological invariants confirm the non-trivial topology: moderately strained \graySn is both a strong TI and a topological crystalline insulator. This necessitates that according to the bulk-boundary correspondence \cite{Hasan2010} a TSS has to bridge (`interconnect') the \GSP and \GSM bands: this state is unambiguously identified as TSS2.

The absence of both TSS1 and TSS2 in the bulk electronic structure [Fig.~\ref{THEOR-1N}(a)] is clear evidence for their surface character. It is also evident from comparing Fig.~\ref{THEOR-1N}(b) with Fig.~\ref{THEOR-1N}(c) that TSS1 has its LDOS maximum at the surface while TSS2 is located well below the surface. The different surface localization is addressed in Fig.~\ref{THEOR-1N}(d). Due to overlap with projected bulk bands, both TSS1 and TSS2 are strictly speaking so-called 'surface resonances', which are characterized by an enhanced LDOS in the surface or subsurface region being degenerate with a finite LDOS in the bulk. Yet TSS1 decays in general exponentially toward the bulk, thereby being more reminiscent of a surface state; in contrast, the LDOS of TSS2 decays and oscillates, which is more typical of a \textit{surface resonance}.

To probe the subsurface-localized TSS2 experimentally we utilized SX-ARPES which has a higher probing depth compared to low-photon-energy ARPES \cite{TPP-2}. Owing to an excellent $k_{\perp}$ resolution, the experimental SX-ARPES data acquired with polarized $\unit[352]{eV}$ and $\unit[684]{eV}$ photons detect mostly states close to the bulk $\mathrm{\Gamma}$ points [Fig.~\ref{ARPES-1}(a)]; they show clearly all three occupied bulk valence bands, i.\,e., \GEP, \GSM, and \GSP [Figs.~\ref{ARPES-1}(b) and ~\ref{ARPES-1}(c)]. 
While at both photon energies the intensity from TSS1 near \EF is very weak, TSS2 located between the surface-projected \GSM and \GSP bulk bands at $\approx \unit[-0.8]{eV}$ is unequivocally resolved [Fig.~\ref{ARPES-1}(e) and ~\ref{ARPES-1}(f)]; it is strikingly similar to TSS2 in the calculations [Fig.~\ref{THEOR-1N}(c)]. The fact that along the surface perpendicular [001] direction TSS2 appears only near bulk $\mathrm{\Gamma}$ points and persists in spectra acquired with high probing depth ($h\nu = \unit[684]{eV}$) clearly indicates its surface resonance character.
The apparent broadening of the \GSP band at $h\nu = \unit[352]{eV}$ is most likely a consequence of a small yet finite $k_{\perp}$ resolution \cite{Strocov2003}.

While TSS1 appears very pronounced between the \GSM and the \GEP bands at photon energies around \unit[20]{eV}, the intensity of TSS1 is small at soft X-ray photon energies (see Ref.~\onlinecite{Suppl2016}). Interestingly, in a study by Liu \textit{et al.} on HgTe(110) \cite{Liu2015b}, which closely resembles the electronic structure of \graySn, a counterpart of TSS1 was observed over a wide SX-photon energy range. The small intensity of TSS1 in \graySn might be better understood by future SX-ARPES photoemission simulations.

As depicted in Fig.~\ref{ARPES-1}(d), the \GEP band crosses \EF, indicating that the film is $p$-doped despite the Te doping; this may be explained by Sn vacancies or diffusion of In atoms from the substrate into the film \cite{Barfuss2013c}. We note that the small negative partial band gap between both \GEP bands described above and in Ref.~\onlinecite{Suppl2016} is likely to occur above \EF and is, for the small strain induced through the InSb substrate ($\unit[0.13]{\%}$), below the resolution limit of our experiment.
\begin{figure} [!ht]
	\includegraphics[clip, trim=1cm 0cm 1.5cm 0cm, width=0.95\linewidth]{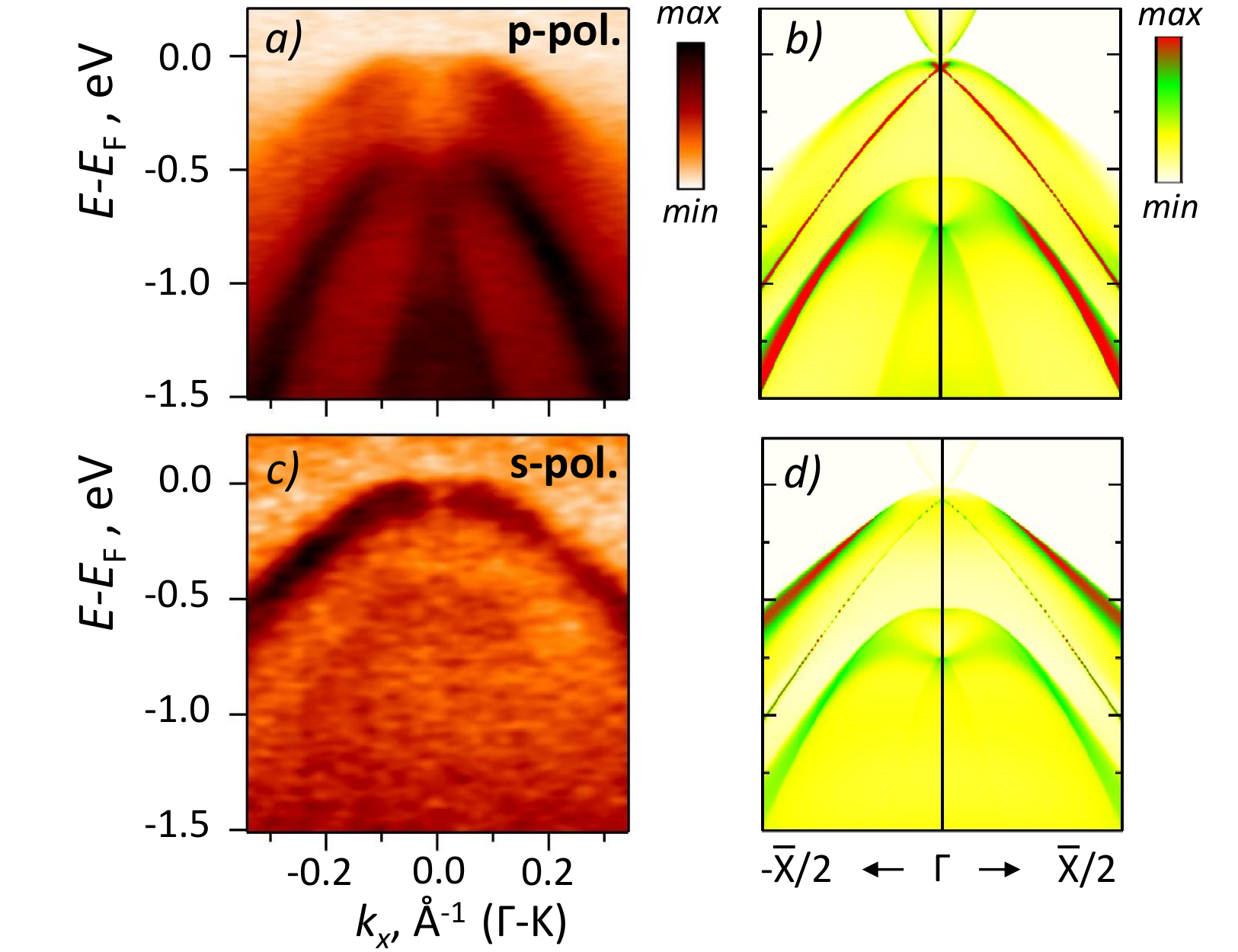}
	\caption{Orbital composition of the electronic states. a) ARPES map $I(E, k_{x}$) measured along $\mathrm{\Gamma K}$ ($\overline{\Gamma}\, \overline{\mathrm{X}}$) for $h\nu = \unit[352]{eV}$ and p-polarized light. The main contribution comes from electronic states that are symmetric w.\,r.\,t.\ the $\mathrm{X \Gamma Z}$ mirror plane. b) Computed spectral density of symmetric orbitals (summed over first six DLs). c) As a) but for s-polarized light. High intensities are due antisymmetric states. d) As b) but for antisymmetric orbitals.}
	\label{ARPES-2}
\end{figure}
In order to better understand the different spatial distribution of the TSS1 and TSS2, we explore their \textit{orbital composition} using two distinct polarizations of the incident light (Fig.~\ref{ARPES-2}). In the experimental geometry, the measurement plane coincides with a $\mathrm{(\overline{1}10)}$ mirror plane ($\mathrm{X \Gamma Z}$). Therefore, for light polarized parallel (even) to the measurement plane (p-polarized light), photoemission from initial states whose wavefunction is antisymmetric (odd) with respect to the $\mathrm{(\overline{1}10)}$  mirror plane is suppressed, since the final state of the photoelectron is even \cite{Hermanson1977}. Thus, only states with dominant $s$, $p_{z}$, and symmetric combinations of $p_{x}$ and $p_{y}$ orbitals are visible in Fig.~\ref{ARPES-2}(a). In turn, with perpendicular polarization (odd or s-polarized light) one detects states with antisymmetric wavefunctions [i.\,e., antisymmetric combinations of $p_{x}$ and $p_{y}$ orbitals; Fig.~\ref{ARPES-2}(c)].
To explain the experimental polarization dependence, we computed orbital-resolved spectral densities; only states symmetric or antisymmetric to the $\mathrm{(\overline{1}10)}$ plane summed over the first six DLs are shown in  Figs.~\ref{ARPES-2}(b) and ~\ref{ARPES-2}(d). The results fit very well to the experimental SX-ARPES data and provide a strong hint towards a similar orbital composition of TSS2 and the projected bulk states.

Further investigations of the orbital compositions show that the projected bulk states between the \GSP and \GSM levels are mainly composed of $p_{x}$ and $p_{y}$ orbitals but lack $p_{z}$ contributions (Fig.~4 in Ref.~\onlinecite{Suppl2016}). The same orbital composition is found for TSS2, which explains the high degree of its hybridization with the bulk states and its pronounced \emph{surface resonance} character. On the contrary, TSS1 is mainly composed of $p_{z}$ orbitals; it therefore hybridizes much weaker with the $p_{x}$ and $p_{y}$ bulk states below \GEP, which results in its exponential decay toward the bulk. Recalling that despite the absence of a global band gap, TSS1 has sharp dispersive peak in ARPES (i.e., long photo-hole lifetime), we conclude that the decay channel of the TSS1 photo-hole through the bulk states is reduced due to their different orbital composition and strong surface localization of TSS1. These results verify an \textit{"orbital protection mechanism"} for such surface resonant TSSs, which can occur in addition to a topological protection from backscattering.

 Bearing resemblance to Weyl or Dirac semimetals in which both topological bulk and surface states cross the Fermi level \cite{Bulmash2014,Inoue2016,Lv2015,Rauch2015,Ruan2016,Yang2014a}, TSSs can appear degenerate with the surface-projected bulk band structure as long as they reside within \textit{local} energy- and wavevector-dependent band gaps \cite{Fukui2007,Teo2008,Thonig2016,Kutnyakhov2016}. Moreover, such states were observed to appear also in spin-orbit induced partial gaps away from the \EF in well-known layered topological insulators, e.g. $ \mathrm{Sb_{2}Te_{3}}$ \cite{Pauly2012,Seibel2015} and $ \mathrm{Bi_{2}Se_{3}}$ \cite{Niesner2012, Sobota2013}. However, a pair of surface-resonance TSSs which share the same bulk band and have very different spatial decay character is so far a unique property of \graySn.
 

In conclusion, the surface and subsurface electronic structure of \graySn thin films strained compressively in the (001) plane is studied theoretically and experimentally. In agreement with \textit{ab initio}-based electronic structure calculations, the bulk-sensitive SX-ARPES data allow to reveal an additional topological surface state located between the projected \GSM and \GSP bands, to the best of our knowledge unreported so far \cite{Barfuss2013c,Ohtsubo2013b,Rojas-Sanchez2016}. The nontrivial topology of this surface state is proven by the topological invariants of the associated bulk states. This new TSS of \graySn in the projected valence bands has a pronounced surface resonance character, in clear contrast to the surface-localized TSS1. 
Based on the analysis of orbital compositions, the different decay characters of TSS1 and TSS2 are attributed to their hybridization with surface-projected bulk states. Our findings thus reveal a new type of TSS in \graySn that is coupled by orbital symmetries to the bulk states in the whole energy range. In addition, we find that an \textit{"orbital protection mechanism"} can effectively protect the surface-resonant TSS against hybridization, as in TSS1 in the present case. Recently, a pair of surface states reminiscent of TSS1 and TSS2 discussed in the present work was also found in half-Heusler compounds whose bulk-band structure is quite similar to those of \graySn and HgTe \cite{Liu2011,Logan2016}. The authors of those studies attribute a nontrivial topology to the surface state near \EF; considering our findings one may speculate that the other surface state is of topological origin, too.

This work was supported by Sonderforschungsbereich SFB 1170 ToCoTronics and Priority Program SPP 1666 of the Deutsche Forschungsgemeinschaft (DFG). A.F. thanks the Deutsche Forschungsgemeinschaft for support (Grant No. FOR 1162) and the J{\"u}lich Supercomputing Centre for providing the computer resources (project No. hwb03).

\bibliographystyle{apsrev4-1.bst}

%

\end{document}